\documentclass[11pt]{article}
\usepackage{amsfonts,amssymb,amsmath}

\newcommand{\ket}[1]{\mbox{$| #1 \rangle$}}
\newcommand{\bra}[1]{\mbox{$\langle #1 |$}}
\newcommand{\braket}[2]{\mbox{$\langle #1 | #2 \rangle$}}

\def\squareforqed{\hbox{\rlap{$\sqcap$}$\sqcup$}}
\def\qed{\ifmmode\squareforqed\else{\unskip\nobreak\hfil
\penalty50\hskip1em\null\nobreak\hfil\squareforqed
\parfillskip=0pt\finalhyphendemerits=0\endgraf}\fi}
\newtheorem{theorem}{Theorem}
\newtheorem{definition}{Definition}
\newtheorem{lemma}[theorem]{Lemma}
\newenvironment{proof}{\begin{trivlist}\item[]{\flushleft\bf Proof }}
{\qed\end{trivlist}}

\DeclareMathSymbol{\leqslant}{\mathrel}{AMSa}{"36}  

\newcommand{\subgroup}{\leqslant}          
   
\newcommand{\complex}{\mathbb{C}}   

\title{Quantum State Detection via Elimination}

\author{Mark Ettinger\,%
\thanks{\,\mbox{NIS--8}, \mbox{MS~B230}, 
Los Alamos National Laboratory, Los Alamos, NM~87545, USA.
Email: \texttt{\boldmath ettinger$\mathchar"40$lanl.gov}.}\\%
{\protect\small\sl LANL\/}%
\and
Peter H{\o}yer\,%
\thanks{\,BRICS, Department of Computer Science, 
University of Aarhus, \mbox{DK--8000} \mbox{{\AA}rhus~C}, Denmark.
Email: \texttt{\boldmath hoyer$\mathchar"40$brics.dk}.}\\%
{\protect\small\sl BRICS\/}}

\date{{\normalsize May 27, 1999}}

\begin{document}

\maketitle

\begin{abstract}
We present the view of quantum algorithms as a search-theoretic problem.
We show that the Fourier transform, used to solve the Abelian hidden
subgroup problem, is an example of an efficient elimination observable which
eliminates a constant fraction of the candidate secret states with high
probability.  Finally, we show that elimination observables do not always
exist by considering the geometry of the
hidden subgroup states of the dihedral group
$D_N$.
\end{abstract}


\section{Introduction}
In the classic game of ``Twenty Questions'', Player~$1$ thinks of a secret 
number between 1 and~$N$.  Player~$2$ tries to guess the number in as few
tries as possible by asking questions of the form, ``Is the secret number
less than or equal to $x$?''  It is well known that if Player $1$ always 
answers correctly then $\lceil\log N \rceil$ questions 
are necessary and sufficient to determine the number.  
Questions like this are studied
in combinatorial search theory~\cite{Aig88,AW87}.  
More formally, a search problem
is a pair $(S,\mathcal{F})$ 
where $S$ is a set called the search space
and $\mathcal{F}$ is a set of functions defined on~$S$,
called the set of allowable questions.
There are two players, and the problem is for Player~$2$ to 
determine a secret element $x_0 \in S$ initially only known by
Player~$1$.
To~learn this secret,  
Player~$2$ can ask questions $f \in \mathcal{F}$
to which Player~$1$ must answer with elements $x \in S$ 
for which $f(x) = f(x_0)$.
Another way of phrasing this is to say that the allowable questions 
form a subset of the partition lattice of $S$ 
and the answer discloses the block of the chosen partition question 
which contains the secret element~$x_0$.

In the present paper we cast known quantum algorithms in a similar light.
We present a quantum search-theoretic game in which Player $1$ chooses a 
secret quantum state $\rho$ from among a set $S$ 
of possible states and supplies
Player $2$ with multiple copies of the secret state.  Player $2$ applies
a sequence of observables for the purpose of discovering the secret state.
The observables are the quantum analogues of the partition questions of 
traditional search theory.
Certain sets $S$ have a geometric quality that permits the construction of a
POVM called an {\em efficient elimination observable\/} which eliminates any 
nonsecret state with high probability.  In 
this case the secret state is efficiently discoverable, i.e., in 
$O(\log^{O(1)}(|S|))$ observables, with high probability.
This is the quantum analogue of Twenty Questions.  We show that the 
Fourier Transform, the well-known solution of the hidden subgroup problem,
is such an observable.  This permits an understanding of the hidden subgroup
solution independent of harmonic analysis.  We then use the hidden subgroup
states of the dihedral group $D_N$ as an example of a set $S$ which does 
not possess the geometric property permitting the construction of an
elimination observable.  This example shows the essential nature of the
probabilistic data discussed in~\cite{EH99}.

\section{Elimination Observables}
We consider the following quantum state detection game played between
two players.  There is a set of possible {\em secret states\/} $S =
\{\rho_1,\dots,\rho_N\}$ known to both players.  Player 1 (who may be
``nature'')  chooses a secret state $\rho_0 \in S$ and provides
multiple copies of the secret state when requested to do so by Player
2.  The task for Player 2 is to design an observable (POVM) or
sequence of observables which allows him to guess the secret state in
as few requests to Player 1 as possible.  Similar scenarios have been
considered~\cite{Hel76}.

\begin{definition}
Let $S = \{\rho_1,\dots,\rho_N\}$ be a set of states and $A =
\{A_1,\dots,A_m\}$ a POVM.   For each $\rho \in S$ we define the 
{\em elimination set of $\rho$ with respect to $A$\/} as $E_A(\rho) =
\{A_i \in A : \textup{tr}(\rho A_i) = 0\}$ 
and the {\em elimination operator of
$\rho$ with respect to $A$\/} as 
$A^\perp_{\rho} = \sum_{A_i \in E_A(\rho)}{A_i}$.  
$A$ is called an {\em efficient elimination observable\/} for
$S$ if there exists a constant $c \in (0,1]$ such that for all $\rho,
\rho^\prime \in S$ we 
have $\textup{tr}(\rho A^\perp_{\rho^\prime}) \geq c$
or $\textup{tr}(\rho^\prime A^\perp_{\rho}) \geq c$.  
An efficient elimination POVM $A$ is called {\em
optimal\/} if for all $A_i \in A$, $\rho \in S$, and positive operators
$B$, if $\textup{support}(B) \subseteq \textup{support}(A_i)$ 
then $\textup{tr}(\rho B) = 0$
implies $\textup{tr}(\rho A_i) = 0$.
\end{definition}

Intuitively, an efficient elimination observable allows us to detect a
secret state in a polynomial number of experiments with high
probability.  Suppose the secret state is $\rho = \rho_0$.  If $A$ is an
efficient elimination observable then for any other $\rho^\prime \in
S$, by measuring $A$ on $\rho$ we either eliminate the possibility that
$\rho^\prime$ is the secret state with probability at least $c$ or
the probability that the secret state is $\rho^\prime$ drops
exponentially.  This
suffices to eliminate $\rho^\prime$ in a polynomial number of
experiments with exponentially high probability.  
Thus in a polynomial number of measurements we eliminate
everything except the secret state $\rho_0$.  

The notion of an optimal efficient elimination observable captures the
notion that it does not help to refine any of the outcomes $A_i \in A$
because to do so would not provide any increased information in the
form of knowing more eliminated states.

For any set of states $S$ we may attempt to construct an 
elimination measurement in the following natural way.  
For each $\rho \in S$ define 
$\rho^\perp = \textup{ker}(\rho) = \textup{support}(\rho)^\perp$ 
to be the orthogonal complement 
of the support of $\rho$ in $\mathcal{H}$.  We now take intersections of
various subsets $\{\rho_1^\perp,\dots,\rho_m^\perp\}$ in the hope of finding 
subspaces of $\mathcal{H}$ which lie in many of the $\rho^\perp$s.  The idea
is that if $L \leq \rho_{i_1}^\perp \cap \cdots \cap \rho_{i_k}^\perp$ 
is such an {\em elimination subspace\/} and
if $L$ is the outcome of a measurement, then we may 
eliminate $\rho_{i_1},\dots,
\rho_{i_k}$ as possibilities for the secret state.  This method does not
always yield an efficient elimination observable as we will see below in the
case of hidden subgroups of the dihedral group $D_N$.  To obtain an 
efficient elimination observable we need to find enough of these elimination
subspaces to span $\mathcal{H}$.  Sometimes the geometry of $S$ does not
permit this.  However in the case of hidden subgroups of finite, Abelian
groups this technique does work and, interestingly enough, 
yields the same observable which has been used in previous quantum algorithms,
namely the Fourier observable.  

\section{Elimination and Hidden Subgroup States}
In the hidden subgroup problem we are given a finite
group $G$ and an oracle function on $G$ which is promised to be
constant and distinct on cosets of some subgroup $H \subgroup G$.  
Let $K$ be a transversal for $H$ in $G$ and let $C = K^m$.  Using
the oracle we may easily prepare {\em multicoset states of order m\/}, 
a tensor product of $m$ coset states of $H$, i.e.{}
$$\ket{\psi(H,c)} = \ket{c_1H} \otimes \cdots \otimes \ket{c_mH},$$ 
where for any non-empty subset $Y \subseteq G$, 
$$\ket{Y} = \frac{1}{\sqrt{|Y|}} \sum_{y \in Y} \ket{y}$$ and 
$c = (c_1,\dots,c_m) \in C$.
Define the mixed state 
$$\rho_H = \left(\frac{|H|}{|G|}\right)^m \sum_{c \in C} 
|\psi(H,c)\rangle \langle|\psi(H,c)|.$$
The state $\rho_H$ is equal to the mixed state
we obtain by applying the oracle function $m$ times 
and tracing out the $m$ registers holding the function-values.

Let $S_G = \{\rho_H : H \subgroup G\}$ be the set of
possible mixed states.
Note that $S_G$ also implicitly
depends on $m$.  In~\cite{EHK99} it shown that these mixed states are
distinguishable for $m$ being on the order of~$\log|G|$.  
In other words, hidden subgroup
states are distinguishable in a small number of oracle calls.  
A~number of problems are reducible to hidden subgroup problems including
discrete logarithm, graph isomorphism, code equivalence, and various
equivalent problems thought to be strictly harder 
than graph isomorphism~\cite{Hof82}.  
As an example of this last category we mention {\em
  restricted graph automorphism\/}, where given a graph $\Gamma$ on $n$
vertices and a subgroup $J$ of $S_n$ given by generators one should find a
set of generators for the subgroup $\textup{Aut}(\Gamma) \cap J$.  

It is well known~\cite{Kitaev95} 
that when $G$ is Abelian one can
efficiently find a hidden
subgroup $H$ using only a single coset state at a time, i.e.{}  working
in the Hilbert space $\complex[G]$, utilizing the quantum Fourier
transform.  The Fourier transform is a change of basis transformation from the
point mass basis to the basis of characters of $G$.  Let 
$\hat{G} = \{\chi_1,\dots,\chi_{|G|}\}$  be the group of characters of $G$ and 
let $\ket{\chi} = \frac{1}{\sqrt{|G|}}
\sum_{g \in G} {\chi(g) \ket{g}}$.  We may 
alternatively refer to the {\em Fourier observable\/} $F(G)$ which is the 
self-adjoint operator defined by the character basis: 
$$F(G)= \sum_{i = 1}^{|G|}{i \ket{\chi_i}\bra{\chi_i}}.$$

The following result casts this well known fact in the
state distinguishability paradigm.

\begin{theorem}
If $S_G$ is the set of hidden subgroup states of an Abelian 
group $G$ then
the Fourier observable is a refinement of the 
unique optimal elimination POVM for $S_G$.
\end{theorem}

This result implies that we may rederive the Fourier observable 
from strictly geometric considerations.  
It is our hope that this geometric perspective may
yield insight into the value of new observables 
for similar state distinguishability problems.

The theorem follows immediately from the following two lemmas.  
Although the proofs of the lemmas rely on results from harmonic
analysis for brevity, we emphasize that one may derive them
without recourse to these results.  
Admittedly this may involve difficult calculations but the
point is that this basic quantum algorithm may be understood 
without any knowledge of Fourier analysis.

The first lemma is basic and describes the elimination subspaces 
of hidden subgroup states.
Let $H \subgroup G$.  Define the orthogonal group to $H$ as 
$H^\perp = \{\chi : \chi(h) = 1 \textup{ for all } h \in H\}.$

\begin{lemma}
For any $H \subgroup G$, $\rho_H^\perp = 
\langle \,\ket{\chi}\, : \chi \not \in H^\perp\rangle$.  
\end{lemma} 

\begin{proof}
Notice $\rho_H$ corresponds to a random choice of a coset state $\ket{cH}$.
By basic results
in Fourier analysis on Abelian groups~\cite{TAL97} we have $|H^\perp| = 
\frac{|G|}{|H|}$ and we may write the coset state as
$$\ket{cH} = \sqrt{\frac{|H|}{|G|}} 
 \sum_{\chi \in H^\perp}\chi(c)\ket{\chi}.$$
Therefore for all $\chi \not \in H^\perp$ we have $\braket{\chi}{cH} = 0$.
This means $\chi \in \rho_H^\perp$ and therefore
$\langle \ket{\chi} :
 \chi \not \in H^\perp\rangle \subseteq \rho_H^\perp$.
For the other inclusion,
notice that for $c,d \in G$ with $c \not\in \ket{dH}$, we
have $\braket{cH}{dH} = 0$.  
This means that $\dim(\rho_H^\perp) 
= \dim(\mathcal{H}) - \frac{|G|}{|H|}$.  
But $|\{\chi : \chi \not \in H^\perp\}| = 
|G| - \frac{|G|}{|H|} = \dim(\mathcal{H}) - \frac{|G|}{|H|}$.  
The lemma follows.
\end{proof}

Before stating
the second lemma we require some definitions which describe the optimal 
elimination observable.

Define an equivalence
relation on $\hat{G}$ by $\chi_1 \equiv \chi_2$ if and only if 
$\chi_1 \in \langle\chi_2\rangle$ and
$\chi_2 \in \langle \chi_1 \rangle$.  Let the equivalence classes be 
$\{[\chi_1],\dots,[\chi_s]\}$.   For each class let 
$\mathcal{H}_i 
= \langle \ket{\chi} : \chi \in [\chi_i] \rangle$  
be the subspace of $\mathcal{H} = \complex[G]$ spanned by the 
members of~$[\chi_i]$.  
Let $P_i$ be the projection onto $\mathcal{H}_i$ and
define the self-adjoint operator $A(G) = \sum_{i=1}^s {i P_i}$.

\begin{lemma}
If $S_G$ is the set of hidden subgroup states of $G$ 
then the unique optimal elimination POVM for $S_G$ is~$A(G)$.
\end{lemma}

\begin{proof}
We first prove $A(G)$ is an efficient elimination observable.  The argument
is actually the standard argument that the quantum algorithm efficiently
finds a hidden subgroup but phrased in the present terminology.  Suppose the
hidden subgroup is $H$, i.e.{}  $\rho = \rho_H$, and let $J \subgroup G$.  
If $J \subgroup H$ then $H^\perp \subgroup J^\perp$, and further, 
if $J$ is strictly contained in $H$, 
then $\frac{|J^\perp|}{|H^\perp|} \leq \frac{1}{2}$.  
Using the formula above we see
$\textup{tr}(\rho_J A(G)_{\rho_H}^\perp) \geq \frac{1}{2}$.
If $J \not \leq H$ 
then $$\frac{|H^\perp|}{|J^\perp \cap H^\perp|} \leq \frac{1}{2}$$ 
and thus $\textup{tr}(\rho_H A(G)_{\rho_J}^\perp) \geq \frac{1}{2}$.  

We now show optimality and uniqueness.  Let $\chi \equiv \chi^\prime$.  Then
note that for any $H$, $\chi \in H^\perp$ 
if and only if $\chi^\prime \in H^\perp$.
So by the first lemma
$\chi \in \rho_H^\perp$ 
if and only if $\chi^\prime \in \rho_H^\perp$.
This means that $A(G)$ is a refinement 
of {\em any\/} elimination observable.
Unique optimality follows from this.
\end{proof}

\section{Subgroup States of $D_N$ Cannot be Eliminated}
In~\cite{EH99} the hidden subgroup problem over $D_N$ is considered.  
The quantum algorithm given in that paper 
results in probabilistic data 
which information-theoretically determines the hidden subgroup but for 
which there is no known processing technique which 
enables the hidden subgroup to be found efficiently.  
Obviously, it is therefore reasonable to seek 
another quantum algorithm based on an elimination observable 
for which any necessary postprocessing may be performed efficiently.  
Unfortunately, there does not exist any such efficient elimination 
observable for the hidden subgroup states of $D_N$, which we now show.

For the sake of clarity let us consider $D_P$ where $P$ is a prime.  Similar
arguments hold when $N$ is composite but are cluttered by irrelevant details.
Notationally we write elements of $D_P$ as ordered pairs 
$(a,b) \in Z_P \rtimes Z_2$.  Without loss of generality we may assume
the hidden subgroup has the form $H = \{(0,0),(k,1)\}$~\cite{EH99} and
thus refer to hidden reflections.  We 
try to construct an elimination observable in the generic way outlined in
the first section.  This method fails however because the intersection
of any two elimination spaces always results in the same one-dimensional
space.  We now make this precise.

Let $\mathcal{H}_1$  be the elimination subspace of the 
hidden reflection $(k_1,1)$.  This means that all vectors in $\mathcal{H}_1$
are orthogonal to all coset states $\ket{(a,0)} + \ket{(a+k_1,1)}$.  It is easy
to see that $\mathcal{H}_1$ consists of the subspace 
of $(k_1,1)$-antiperiodic vectors, i.e., vectors of the form
$${\sum_{i=0}^{P-1} \lambda_i \ket{(i,0)}}
  -{\sum_{i=0}^{P-1} \lambda_i \ket{((i+ k_1) \text{ mod } P,1)}},$$  
where $\lambda_0,\ldots,\lambda_{P-1} \in \complex$ are complex numbers.
Similarly
the elimination space $\mathcal{H}_2$ 
of the hidden reflection $(k_2,1)$
contains precisely the $(k_2,1)$-antiperiodic vectors.  
So $\mathcal{H}_1 \cap
\mathcal{H}_2$ contains those vectors which are both $(k_1,1)$-antiperiodic 
and $(k_2,1)$-antiperiodic.  But because $P$ is prime the only vectors of this
form lie in the one-dimensional subspace spanned by the vector
$$v = \sum_{i=0}^{P-1} \ket{(i,0)} -
   \sum_{i=0}^{P-1} \ket{(i,1)}.$$  
Therefore the intersection of any two (or more) 
elimination spaces is the space spanned by~$v$.  Clearly $\complex[D_P]$ is
not spanned by subspaces which satisfy the elimination observable criterion.
This shows the nonexistence of an elimination observable for~$D_P$.

\section{Conclusion}
We have shown that the quantum algorithm for finding hidden subgroups of
finite Abelian groups is an example of a quantum state distinguishability
game.  This game is a search-theoretic quantum analogue of 
``Twenty Questions''.  The Fourier transform is the unique optimal
elimination observable corresponding to the set of hidden subgroup states.
We have
also shown an example of a search space with a geometry that prevents
the construction of an elimination observable, the hidden subgroup states of
$D_N$.  We mention several possibilities for future work.  The classical
search-theoretic literature seems to be concerned exclusively with 
elimination search.  What types of probabilistic search are possible and
can they be adapted to situations like $D_N$ where elimination is impossible?
Finally, are there other, natural problems
that may be 
phrased as state distinguishability problems and solved via elimination
observables?

\end{document}